\documentclass[aip, pof, preprint]{revtex4-1}

\usepackage{graphicx} 
\usepackage{dcolumn}
\usepackage{bm}
\usepackage{amssymb}
\usepackage{amsmath}
\usepackage{wasysym}
\usepackage{color}
\usepackage{hyperref}
\usepackage{dcolumn}
\usepackage{float}
\usepackage{flushend}
\usepackage{multirow}
\usepackage{cancel}
\hypersetup{
colorlinks,
citecolor=blue,
filecolor=blue,
linkcolor=blue,
urlcolor=blue}

\newcommand{\be}{\begin{equation}}
\newcommand{\ee}{\end{equation}} 
\newcommand{\bea}{\begin{eqnarray}}
\newcommand{\eea}{\end{eqnarray}}

\begin{document}

\title{ENERGY SPECTRA AND FLUXES IN DISSIPATION RANGE OF TURBULENT AND LAMINAR FLOWS}

\author{Mahendra K. Verma}
\email{mkv@iitk.ac.in}
\affiliation{Department of Physics, Indian Institute of Technology, Kanpur 208016, India}
\author{Abhishek Kumar}
\email{abhishek.kir@gmail.com}
\affiliation{Department of Physics, Indian Institute of Technology, Kanpur 208016, India}

\author{Praveen~Kumar}
\email{praveenkumar.hcdu@gmail.com}
\altaffiliation[Present address: ]{BARC, Mumbai}
\affiliation{Department of Physics, Indian Institute of Technology, Kanpur 208016, India}
\author{Satyajit Barman}
\email{sbarman@iitk.ac.in}
\affiliation{Department of Physics, Indian Institute of Technology, Kanpur 208016, India}
\author{Anando G. Chatterjee}
\email{anandogc@iitk.ac.in}
\affiliation{Department of Physics, Indian Institute of Technology, Kanpur 208016, India}
\author{Ravi Samtaney}
\email{ravi.samtaney@kaust.edu.sa}
\affiliation{Mechanical Engineering, Division of Physical Sciences and Engineering, King Abdullah University of Science and Technology - Thuwal 23955-6900, Kingdom of Saudi Arabia}
\author{Rodion Stepanov}
\email{rodion@icmm.ru}
\affiliation{Institute of Continuous Media Mechanics, Korolyov 1, Perm 614013, Russia}

\begin{abstract}
Two well-known turbulence models that describe the energy spectrum in the inertial and dissipative ranges simultaneously are by Pao~(1965) and Pope~(2000). In this paper, we compute the energy spectrum $E(k)$ and energy flux $\Pi(k)$ using direct numerical simulations on grids up to $4096^3$, and show consistency between the numerical results and the predictions by the aforementioned models. We also  construct a model for laminar flows that predicts $E(k)\sim k^{-1} \exp(-k)$ and $\Pi(k)\sim  k \exp(-k)$.  Our model predictions match with the numerical results. We emphasize  differences on the energy transfers in the two flows---they are {\em local} in the turbulent flows, and {\em nonlocal} in  laminar flows.
\end{abstract}

\maketitle

\section{Introduction}
\label{sec:intro}

Turbulence is a classic problem with many unresolved issues.  The most well-known phenomenology of turbulence is by \citet{Kolmogorov:DANS1941Structure},  according to which the energy supplied at the large scales  cascades to small scales.  The wavenumber band dominated by the forcing is called {\em forcing range}, while that dominated by dissipation is called {\em dissipative range}. The wavenumber band between these two ranges is termed as {\em inertial range}.  According to \citet{Kolmogorov:DANS1941Structure}, the energy cascade rate or the spectral energy flux is constant in the inertial range.  Quantitatively, the one-dimensional energy spectrum $E(k)$ and the energy flux $\Pi(k)$ in the inertial range are
\bea
E(k) & =  &K_\mathrm{Ko} \epsilon^{2/3} k^{-5/3},  \label{eq:Kolm_Ek} \\
\Pi(k) & = & \epsilon, \label{eq:Kolm_Pik}
\eea
where $\epsilon$ is the energy dissipation rate, and $K_\mathrm{Ko} $ is  {\em Kolmogorov's constant}. The above law has been verified using experiments and high-resolution simulations (see \citet{Frisch:book}, \citet{Mccomb:book:Turbulence}, \citet{Davidson:book:Turbulence}, \citet*{Ishihara:ARFM2009} and references therein).  There is, however, a small correction to the exponent ``-5/3'' due to  intermittency~\cite{Frisch:book}. This issue however is beyond the scope of this paper.  Kolmogorov's theory of turbulence and its ramifications have been discussed in detail in several books~\cite{Leslie:book,Mccomb:book:Turbulence,Frisch:book,Davidson:book:Turbulence,Lesieur:book:Turbulence,Pope:book}.

The energy spectrum of Eq.~(\ref{eq:Kolm_Ek}) is universal, i.e., it is independent of fluid properties, forcing and dissipative mechanisms, etc.  Equation~(\ref{eq:Kolm_Ek})  had been extended to  the dissipative range in the following manner:
\be
E(k)  = K_\mathrm{Ko} \epsilon^{2/3} k^{-5/3} f( k/k_d),
\label{Eq:Ek_all}
\ee
where $f(k/k_d)$ is a universal function, and
\be
k_d = (\epsilon/\nu^3)^{1/4}
\label{eq:kd}
\ee
is the dissipation wavenumber scale, also called Kolmogorov's wavenumber. Pao~\cite{Pao:PF1965}, Pope~\cite{Pope:book}, and Mart\'{i}nez {\em et al.}~\cite{Martinez:JPP1997}  modelled $f(k/k_d)$;  Pao~\cite{Pao:PF1965} proposed that $f(x) \sim \exp(-x^{4/3})$, but according to Pope~\cite{Pope:book}
\be
f(x) \sim  \exp \left\{ -\mathbf{\beta}([ {x}^4 + c_\nu^4 ]^{1/4}   - c_\nu) \right\},
\ee
where $\beta$ and $c_\nu$ are constants. Pope's model~\cite{Pope:book} is in good agreement with earlier experimental results (see Saddoogchi  and Veeravalli~\cite{Saddoughi:JFM1994} and references therein).  Pao~\cite{Pao:PF1965} argued that his predictions fit well with the experimental results of Grant {\em et al.}~\cite{Grant:JFM1962}.  Mart\'{i}nez {\em et al.}~\cite{Martinez:JPP1997} proposed that
\be
E(k)  \sim (k/k_d)^\alpha \exp[-\beta (k/k_d)], \label{eq:Martnez}
\ee
and found good agreement between their predictions and numerical results for moderate Reynolds numbers.

There are only a few numerical simulations that have investigated the dissipative spectrum of a turbulent flow. For example,  Mart\'{i}nez {\em et al.}~\cite{Martinez:JPP1997} computed $E(k)$ for flows with moderate Reynolds number ($\mathrm{Re}$) and showed it to be consistent with the model of Eq.~(\ref{eq:Martnez}). On the other hand, Ishihara {\em et al.}~\cite{Ishihara:JPSJ2005} showed that near the dissipation range, Eq.~(\ref{eq:Martnez}) is a good approximation also for high Reynolds number flows. The energy flux in the dissipative regime of a turbulent flow has not been investigated in detail, either by numerical simulation or experiments.  Note that $\Pi(k)$ in the dissipative range is  assumed to be small and rapidly decreasing, and it is typically ignored. In the present paper, we perform turbulence simulations on very high-resolution grids (up to $4096^3$), and compare the numerically-computed $E(k)$ and $\Pi(k)$  with those predicted by various models.

Laminar flows are ubiquitous in nature; some examples of such flows are---blood flow in arteries, micro and nano fluidics~\cite{Lautrup:book}, mantle convection inside the Earth~\cite{Verma:NJP2017}, laminar dynamo, and passive scalar turbulence with large Schmidt number~\cite{Gotoh:book_chapter:passive_scalar}.  Hence, models of laminar  flows are very useful.  In this paper, we construct a spectral model for such flows. We consider laminar flows with Reynolds number of the order of unity.  For such flows, the nonlinearity and the energy flux are  quite small.  {Researchers have studied the kinetic energy spectrum for such flows and predicted this to be of the form $\exp(-k^2)$ or $\exp(-k)$~\cite{Batchelor:JFM1959_smallSc,Kraichnan:PF1968Scalar,Martinez:JPP1997,Linkmann:PRL2015,Verma:ROPP2017}. }  For example,  see Eq.~(\ref{eq:Martnez}) proposed by \cite{Martinez:JPP1997}.  In the present paper, we show that $E(k) \sim k^{-1}\exp(-k/\bar{k}_d)$ and $\Pi(k) \sim k  \exp(-k/\bar{k}_d)$,
where $\bar{k}_d$ is a measure of dissipation wavenumber, describe the energy spectrum and flux of laminar flows.  We also performed direct numerical simulations (DNS) for  $\mathrm{Re}$ ranging from $17.6$ to $49$ and verified our model predictions with numerical data.

According to Kolmogorov's theory of turbulence, the energy transfers among the inertial-range wavenumber shells are forward and local~\cite{Domaradzki:PF1990,Zhou:PF1993,Verma:Pramana2005S2S}.  That is, a wavenumber  shell (say $m$) transfers maximal energy to its nearest forward neighbour shell ($m+1$), and receives maximal energy from its previous neighbour shell ($m-1$).  The aforementioned phenomena have been verified using numerical simulations. However, there has not been a definitive shell-to-shell energy transfer computation for the dissipative regime of a turbulent flow.  Similar computation for laminar flows is also lacking.  In this paper, we show that the shell-to-shell energy transfers in the dissipative regime of a turbulent flow are local.  However, these transfers for laminar flows are nonlocal.

There are a number of analytical works for modeling turbulence, namely quasi-normal approximation,   eddy-damped quasi-normal Markovian ~\cite{Orszag:JFM1970}, direct-interaction approximation ~\cite{Kraichnan:JFM1959}, etc. These sophisticated models attempt to compute  higher-order correlations and related quantities using various closure schemes~\cite{Frisch:book,Leslie:book}.  The focus of the present paper is on the validation of some of the popular spectral models that predict the energy spectrum and fluxes.  We remark that such studies are very important for applications encountered by engineers, and geo-, astro-, and atmospheric physicists.

The outline of the paper is as follows: In Sec.~\ref{sec:model}, we describe the turbulence models of Pao and Pope, as well as our model for laminar flows.  Sec.~\ref{sec:numerical_results} contains numerical results of high-resolution  simulations of turbulent and laminar flows.  We conclude in Sec.~\ref{sec:conclusions}.

\section{Model description}
\label{sec:model}
An incompressible fluid flow  is described by the Navier--Stokes equations:
\begin{eqnarray}
  \frac{\partial {\bf u}}{\partial t} + ({\bf u} \cdot \nabla) {\bf u}
  & = &  - \frac{1}{\rho} \nabla {p} + {\nu} \nabla^2 {\bf u} + {\bf f}, \label{eq:NS} \\
   \nabla \cdot {\bf u} & = & 0, \label{eq:continuity}
\end{eqnarray}
where {\bf u} is the velocity field, $p$ is the pressure field, $\nu$ is  the kinematic viscosity, and {\bf f}  is the force field.  We take the density $\rho$ to be a constant, equal to unity.  In Fourier space the above equations are transformed to (e.g. see ~\citet{Lesieur:book:Turbulence})
\begin{eqnarray}
\left(\frac{\partial}{\partial t} + \nu k^2 \right) {\bf u}({\bf k}, t ) &  = &  - i {\bf k} {p} ({\bf k},t) -i \sum_{{\bf k=p+q}} {\bf k} \cdot {\bf u}({\bf q}) {\bf u} ({\bf p})   +  {\bf f}({\bf k}) , \label{eq:NSk}\\
 {\bf k} \cdot {\bf u}({\bf k})  & = & 0, \label{eq:continuity2}
\end{eqnarray}
where ${\bf {u}(k)}$, ${p} ({\bf k})$, and ${\bf f} ({\bf k})$ are the Fourier transforms of ${\bf u}$, $p$, and ${\bf f}$ respectively.   The above equations yield the following equation for one-dimensional energy spectrum $E(k,t)$~\cite{Lesieur:book:Turbulence}:
\begin{equation}
\frac{\partial E(k,t)}{\partial t} = T(k,t) - 2 \nu k^2 E(k,t) + \mathcal{F}(k,t), \label{eq:energy}
\end{equation}
where  $T(k,t)$ is the energy transfer to the wavenumber shell $k$ due to nonlinearity, $ \mathcal{F}(k,t)$ is the energy feed by the force, and $-2\nu k^2 E(k,t)$ is the dissipation spectrum. Note that $T(k,t) = -\partial\Pi(k,t)/\partial k$.
In Kolmogorov's model, the energy supply rate $ \mathcal{F}(k,t)$ is active at large length scales (for $k \approx k_f$, where $k_f$ is the forcing wavenumber), and it is absent in the inertial and dissipative range. In a statistically steady state,   $\partial E(k,t)/ \partial t = 0$.
Hence the energy flux $\Pi(k)$ varies with $k$ as
\begin{equation}
\frac{d}{dk}  \Pi(k) = -2 \nu k^2 E(k).
\label{eq:dPibydk}
\end{equation}
The turbulence model of Pao~\cite{Pao:PF1965} is based on the above equation.

In hydrodynamics, it is  convenient to work with dimensionless quantities.  We nondimensionalize $k$ using the Kolmogorov wavenumber $k_d$, which is defined by (\ref{eq:kd}).
The energy flux $\Pi(k)$ is nondimensionalized using $\epsilon$.  Hence,  
\begin{eqnarray}
 \tilde k & = & \frac{k}{k_d}, \\
\tilde \Pi (\tilde k) & = & \frac{\Pi(k)}{\epsilon}. \label{eq:Pi_tilde_k}
\end{eqnarray}
We compensate  Eq.~(\ref{Eq:Ek_all}) such that
\be
\tilde{E}(\tilde{k})  =  \frac{E(k)}{K_\mathrm{Ko} \epsilon^{2/3} k^{-5/3}} =   f_{\nu}(\tilde k).  \label{eq:E_tilde_k}
\ee
Substitution of the above variables in Eq.~(\ref{eq:dPibydk}) yields
\begin{equation}
\frac{d}{d \tilde k}  \tilde \Pi(\tilde{k}) = - 2 K_\mathrm{Ko} {\tilde k}^{1/3} f_{\nu}(\tilde k).
\label{eq:nondim_dP_dk}
\end{equation}

After the above introduction, we describe some of the important spectral models for turbulent and laminar flows.

\subsection{Pao's model of turbulent flow}
\label{sec:Pao}

In this paper, we discuss the models of Pao~\cite{Pao:PF1965} and Pope~\cite{Pope:book}.  To contrast the two models, we label the energy fluxes and spectra of these models differently.  For Pao's model, we label the energy spectrum, flux,  nondimensional dissipative function as $\tilde{E}^{(1)}(\tilde{k}) $,  $ \tilde {\Pi}^{(1)}(\tilde k)$ and $ f^{(1)}_{\nu}(\tilde k)$ respectively;   for Pope's model~\cite{Pope:book} we use $\tilde{E}^{(2)}(\tilde{k}) $,  $ \tilde {\Pi}^{(2)}(\tilde k)$ and $ f^{(2)}_{\nu}(\tilde k)$ for the corresponding quantities.

In Eq.~(\ref{eq:nondim_dP_dk}), $\tilde \Pi(\tilde k)$ and $f_{\nu}(\tilde k)$ are two unknown functions.    Hence, to close the problem, Pao~\cite{Pao:PF1965} assumed that the ratio $\Pi(k)/E(k)$ is independent of $\nu$.  Using dimensional analysis, the ratio can be expressed using $\epsilon$ and $k$ as 

\be
\frac{\Pi(k)}{E(k)} = \alpha^{-1} \epsilon^{1/3} k^{5/3},
\ee
where coefficient $\alpha$ is chosen as $K_\mathrm{Ko}$. In terms of the nondimensional variables,
\be
   \tilde{\Pi}^{(1)}(\tilde k)  =  \tilde{E}^{(1)}(\tilde{k})= f_{\nu}^{(1)}(\tilde k).
 \ee
In other words, the dissipative functions for both $\tilde{E}(k)$ and $\tilde{\Pi}(k)$
should be of the same form.  Thus  Eq.~(\ref{eq:nondim_dP_dk}) yields
\begin{eqnarray}
 f^{(1)}_{\nu}(\tilde k) & =  & \exp{\left(- \frac{3}{2} K_\mathrm{Ko} \tilde k^{4/3}\right)}. \label{eq:Pao_fetak}
\end{eqnarray}

\subsection{Pope's model of turbulent flow}
\label{sec:Pope}

Pope~\cite{Pope:book} constructed another popular model for the turbulent flow.  For this model, we  denote the energy spectrum, flux, nondimensional dissipative  function $f_{\nu}(\tilde k)$ as $\tilde{E}^{(2)}(\tilde{k}) $,  $ \tilde{\Pi}^{(2)}(\tilde k)$ and $ f^{(2)}_{\nu}(\tilde k)$ respectively.   Pope~\cite{Pope:book} proposed that
\begin{equation}
E^{(2)}(k) = K_\mathrm{Ko} \epsilon^{2/3} k^{-5/3}  f_L(k L) f^{(2)}_\nu(k /k_d)
\label{eq:Ek_Pope}
\end{equation}
with the functions $f_L(kL)$ and $f_\nu(k /k_d)$ specifying the  large-scale and dissipative-scale deviations from ``-5/3'' power law, respectively:
\begin{eqnarray}
f_L(kL) & = & \left( \frac{kL}{[(kL)^2 + c_L]^{1/2}} \right)^{5/3+p_0},
\label{eq:fL} \\
f^{(2)}_\nu(\tilde{k}) & = & \exp \left[ -\beta \left\{ [ {\tilde{k}}^4 + c_\nu^4 ]^{1/4}   - c_\nu \right\} \right],
\label{eq:feta}
\end{eqnarray}
where the $c_L, c_\nu, p_0, \beta$ are constants. Since we focus on the inertial and dissipative ranges, for which $k \gg 1/L$,
we set $f_L(kL) = 1$. In the high Reynolds number limit, $c_\nu \approx 0.47 \beta^{1/3}/K_\mathrm{Ko}$~\cite{Pope:book}. We choose $\beta = 5.2$ as prescribed  by Pope~\cite{Pope:book}. Substitution of  $\tilde{E}^{(2)}(\tilde{k}) $ and  $ \tilde{\Pi}^{(2)}(\tilde k)$ in Eq.~(\ref{eq:nondim_dP_dk}) yields the following solution
 \begin{equation}
\tilde{\Pi}^{(2)}(\tilde k) = \tilde{\Pi}^{(2)}(\tilde k_0)   - 2 K_\mathrm{Ko} \int_{\tilde{k_o}}^{\tilde{k}} {\tilde k}^{'1/3} f^{(2)}_{\nu}(\tilde k') d\tilde{k'},
\label{eq:Pope_Pik}
\end{equation}
which is solved numerically with  $f^{(2)}_\nu(\tilde{k})$ of Eq.~(\ref{eq:feta}).   We set $\tilde{\Pi}^{(2)}(\tilde k_0) = 1$ at small $k_0$.   Refer to Pope~\cite{Pope:book} for the detailed derivation of above relation.

\subsection{Model for the laminar flows}
\label{sec:model_viscous}

We will show later that the models of Pao~\cite{Pao:PF1965} and Pope~\cite{Pope:book} do not provide satisfactory description of laminar flows with $\mathrm{Re} \approx O(1)$.  Earlier Mart\'{i}nez {\em et al.}~\cite{Martinez:JPP1997} had proposed that the energy spectrum in the dissipative regime has  the following  empirical form:
\be
E(k)  \sim (k/k_d)^\alpha \exp[-\beta (k/k_d)],
\label{eq:Ek_Martinez}
\ee
where $\alpha$ and $ \beta$ are constants.
Based on the above form, we propose a new model for the energy spectrum of laminar flows, according to which
\begin{equation}
E(k) = u^2_{\mathrm{rms}} k^{-1}  f_L(k/\bar{k}_d) \exp(-k/\bar{k}_d),
\label{eq:Ek_laminar}
\end{equation}
where $u_{\mathrm{rms}}$ is the rms velocity of the flow, and
\be
\bar{k}_d = \frac{\sqrt{\mathrm{Re}}}{1.3L},
\ee
 with $L$ as the box size, and $\mathrm{Re}=u_{\mathrm{rms}} L/\nu$ is the Reynolds number. Note that $\bar{k}_d$ of a laminar flow differs from $k_d$, the Kolmogorov's wavenumber defined in Eq.~(\ref{eq:kd}).  Interestingly,  the above form of $\bar{k}_d$ follows from
 \be
 \tilde{k}_d = \left( \frac{\epsilon}{\nu^3} \right)^{1/4} \sim  \left( \frac{\nu U^2}{L^2 \nu^3} \right)^{1/4} \sim \frac{\sqrt{\mathrm{Re}}}{L}.
 \ee
Substitution of Eq.~(\ref{eq:Ek_laminar}) in the expression for the energy dissipation rate yields
\bea
\epsilon & = &  \int_0^\infty 2 \nu k^2 E(k)dk  \nonumber \\
& = & 2 \nu  u^2_{\mathrm{rms}} \bar{k}_d^2 \int_0^\infty \bar{k} f_L(\bar{k}) \exp(-\bar{k} ) d\bar{k}  \nonumber \\
& =  & 2 \nu u^2_{\mathrm{rms}} \bar{k}^2_d A,
\label{eq:epsilon_laminar}
\eea
where $\bar{k} = k/ \bar{k}_d$, and $A = \int_0^\infty \bar{k}  f_L(\bar{k}) \exp(-\bar{k} ) d\bar{k}$ is a nondimensional constant.  From numerical simulation we  observe that $A \sim f_L(\bar{k})  \sim \mathrm{Re}^{-3}$.
Hence, for the laminar regime, we nondimensionalize energy spectrum and energy flux as follows:
\begin{eqnarray}
\bar{E}(\bar{k}) & = & \frac{E(k) k \mathrm{Re}^3 }{ u_\mathrm{rms}^2}, \label{eq:Ek_nondim_laminar} \\
\bar \Pi (\bar k) & = & \frac{\Pi(k)}{\epsilon} =  \frac{\Pi(k)}{2\nu A  u_\mathrm{rms}^2 \bar{k}_d^2}.  \label{eq:Pik_nondim_laminar}
\end{eqnarray}
Substitution of Eq.~(\ref{eq:Ek_laminar}) in Eq.~(\ref{eq:dPibydk}) yields
\begin{equation}
\frac{d \bar \Pi(\bar k)}{d \bar k} = - \frac{\bar{k}}{A}  f_L(\bar{k})\exp{(- \bar k)},
\end{equation}
whose solution is
\begin{equation}
\bar{\Pi}(\bar k) = \bar{\Pi}(\bar k_0) - \frac{1}{A} \int_{\bar{k_0}}^{\bar{k}} \bar{k}f_L(\bar{k})\exp(-\bar k)d \bar k,
\label{eq:flux_viscous}
\end{equation}
where $\bar{k}_0$ is the reference  wavenumber.
Note that for $k> k_f$, the dimensionless energy spectrum and flux are of the following form:
\bea
\bar{E}(\bar{k}) & = & \exp(-\bar{k}) \label{eq:model_laminar_Ek}, \\
\bar \Pi (\bar k) & = & (1+\bar{k}) \exp(-\bar{k}) \label{eq:model_laminar_Pik}.
\eea
In the next section we verify the above models using numerical simulations.

\section{Numerical Validation of the Models}
\label{sec:numerical_results}

We perform  direct numerical simulation of  incompressible Navier-Stokes equation [see Eqs.~(\ref{eq:NS}, \ref{eq:continuity})] in turbulent and laminar regimes, and compute the energy spectra and fluxes for various cases.     We employ pseudo-spectral code Tarang~\cite{Verma:Pramana2013tarang} for our simulations.   We use the fourth-order Runge-Kutta scheme for time advancement with  variable $\Delta t$, which is chosen using the CFL condition. {The pseudo-spectral method produces aliasing error, which is overcome by setting (1/3)rd  of the Fourier modes to zero.  This   dealiasing procedure is referred to as ``2/3'' rule~\cite{Canuto:book:SpectralFluid}.}

We compute the energy spectra and fluxes for all the numerical runs during the steady state. {The energy spectrum $E(k)$ is computed using~\cite{Stepanov:PRE2014}
\begin{equation}
E(k) =\frac{4\pi}{M} \sum_{k-1<k^\prime \leq k} \frac{1}{2}| {\bf u}({\bf k}^\prime)|^2 |{\bf k}^\prime|^2,
\end{equation}
where $M$ is the number of modes in the shell between wavenumbers  $k-1$ and $k$. Note that the above formula reduces bias in the energy spectrum at low wavenumbers~\cite{Stepanov:PRE2014}.} The energy flux $\Pi(k_0)$, the rate of  kinetic energy  emanating from the wavenumber sphere of radius $k_0$, is computed using the following formula~\cite{Dar:PD2001,Verma:PR2004}:
\begin{eqnarray}
\Pi(k_0)  =   \sum_{k > k_0} \sum_{p\leq k_0} \mathrm{Im}\{[{\bf k \cdot u(k-p)}]  [{\bf u^*(k) \cdot u(p)}]\}. \label{eq:ke_flux}
\end{eqnarray}

\subsection{Turbulent Flow}
\label{subsec:numerical_turbulent}

We perform our turbulence simulations on $512^3$, $1024^3$, and $4096^3$ grids.  We employ  periodic boundary conditions on all sides of a cubic box of size $(2\pi)^3$.  To obtain a steady turbulent flow, we apply random forcing~\cite{Reddy:PF2014} in the wavenumber band $2\leq k \leq 4$ for $1024^3$ and $4096^3$ grids, but in the band $1\leq k \leq 3$ for $512^3$ grid. We choose a random  initial condition for the $512^3$-grid simulation.  The steady-state data of $512^3$ was used as an initial condition for the $1024^3$-grid run, whose steady-state data is used for $4096^3$-grid simulation.  In all the three cases, the velocity field at the small scales are  well resolved because $k_{\mathrm{max}} \eta $ is  always greater than $1.5$, where $k_{\mathrm{max}}$ is the highest wavenumber represented by the grid points, and $\eta \sim 1/k_d$ is the Kolmogorov's length.       The Reynolds numbers for the $512^3$, $1024^3$, and $4096^3$ grid simulations are $5.7 \times 10^3$, $1.4 \times 10^4$, and $6.8 \times 10^4$ respectively. {We observe that  the energy flux in the inertial range,  the energy {dissipation} rate, and the energy supply rate by the forcing are equal to each other within 2-4\%.  The energy supply rate is chosen as 0.1, but the energy dissipation rate, as well as the energy flux, vary from 0.096 to 0.102.}   The parameters of our  runs for turbulent flows are listed in Table~\ref{table:simulation_details_turbulent}. In the Table, we  report the value of $\epsilon/(u_\mathrm{rms}^3/L)$ which is approximately unity for all three simulations.

\setlength{\tabcolsep}{14pt}
\begin{table}[htbp]
\begin{center}
\caption{Parameters of our direct numerical simulations (DNS) for turbulent flow: grid resolution; kinematic viscosity $\nu$, Reynolds number $\mathrm{Re}$, Kolmogorov constant $K_\mathrm{Ko} $, Kolmogorov wavenumber $k_d$, $k_{max}\eta$, and $\epsilon/(u_\mathrm{rms}^3/L)$. }
\hspace{20 mm}
\begin{tabular}{c  c c  c  c c c}
\hline \hline
Grid &$\nu$ & $\mathrm{Re}$ & $K_\mathrm{Ko}$ & $k_d$   &  $k_{max} \eta$ & $\epsilon/(u_\mathrm{rms}^3/L)$\\[1 mm]
\hline
$512^3$ & $10^{-3}$ & $5.7 \times 10^3$ & $2.2 \pm 0.2$  &$9.8 \times 10^{1}$ & $2.5$ & $0.9$\\
$1024^3$ & $4 \times 10^{-4}$ & $1.4 \times 10^4$ & $1.85 \pm 0.05$  &$2.1 \times 10^{2}$ & $2.4$ & $1.0$\\
$4096^3$ & $8 \times 10^{-5}$ & $6.8 \times 10^4$ & $1.75 \pm 0.05$  &$6.6 \times 10^{2}$ & $3.1$ & $1.0$\\
\hline

\end{tabular}
\label{table:simulation_details_turbulent}
\end{center}
\end{table}

\begin{figure}[htbp]
\begin{center}
\includegraphics[scale = 0.8]{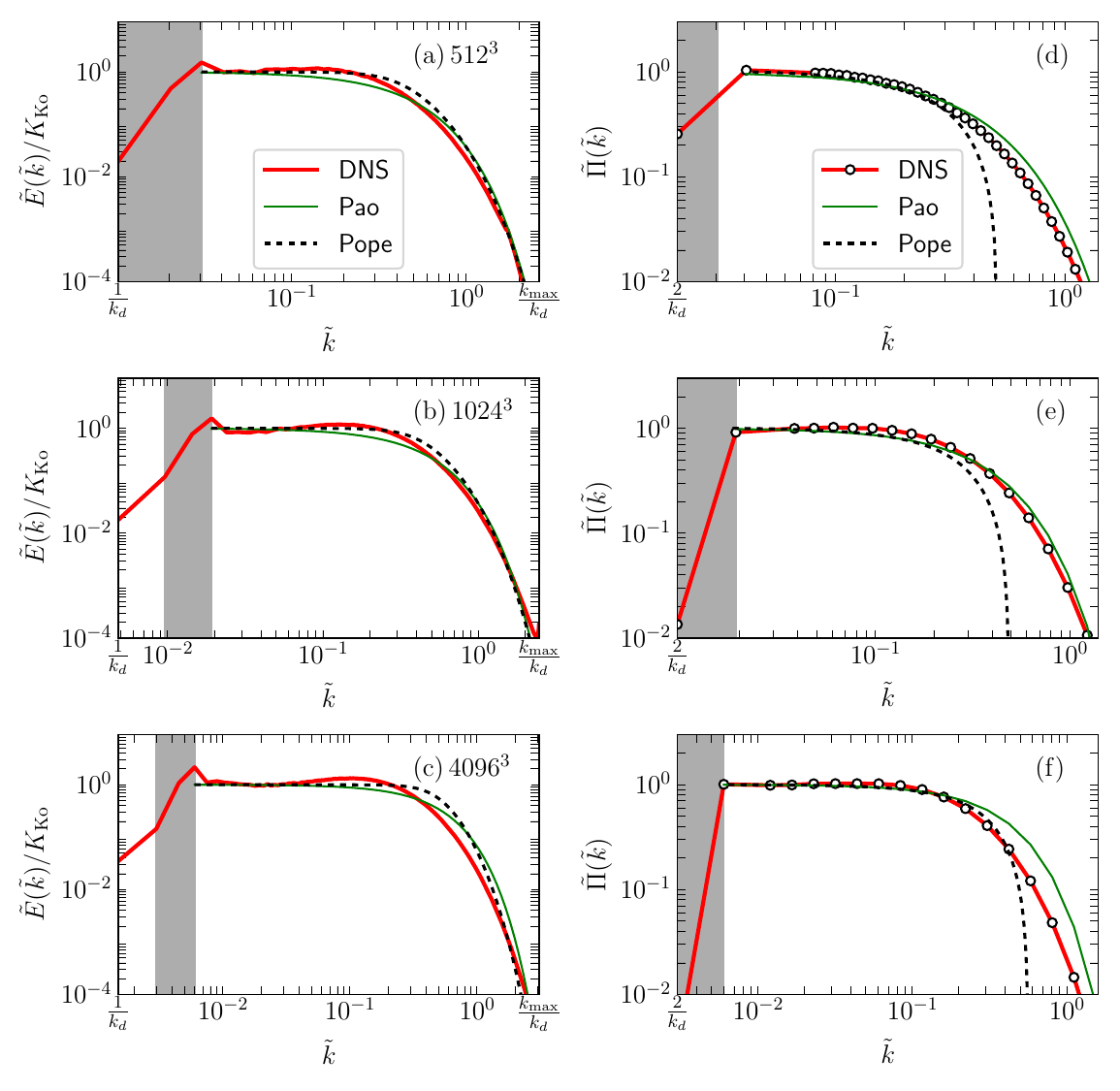}
\end{center}
\setlength{\abovecaptionskip}{0pt}
\caption{For the grid resolutions of $512^3$, $1024^3$, and $4096^3$: (a,b,c) Plots of the normalized energy spectrum $\tilde{E}(\tilde{k})$ vs.~$\tilde{k}$; (d,e,f) plots of normalized energy flux $\tilde{\Pi}(\tilde{k})$ vs.~$\tilde{k}$. See Eqs.~(\ref{eq:Pi_tilde_k}, \ref{eq:E_tilde_k})  for definitions. The plots include the spectra and fluxes computed using numerical data (thick solid line), and the model predictions by Pao (thin solid line) and Pope (dashed line). {The grey  regions indicate the forcing range.}}
\label{fig:spectrum_turb}
\end{figure}

Figure~\ref{fig:spectrum_turb}(a, b, c) exhibits the normalized spectra $\tilde{E}(\tilde{k})$ for  the $512^3$, $1024^3$, and $4096^3$ grid simulations.  Note that the grey  regions in the figures denote the forcing band.  The plots show that the numerical $\tilde{E}(\tilde{k})$  are close to the predictions of both Pao's and Pope's models.     We also compute the Kolmogorov's constant $K_\mathrm{Ko}$ using
\be
K_\mathrm{Ko}  = \epsilon^{- 2/3}\langle{E(k) k^{5/3}}\rangle,
\ee
where an average is taken over scales in the inertial range just after forcing scale.
As shown in Table~\ref{table:simulation_details_turbulent}, the values of $K_\mathrm{Ko}$  varies from {$1.75$ to $2.2$} with errors  in the range of {3\% to 9\%}.  These values are in the same range as  those  reported earlier~\cite{Sreenivasan:PF1995,Yeung:PRE1997,Gotoh:PF2002,Yokokawa:CP2002,Mininni:PRE2008,Donzis:JFM2010Bottleneck}.  The numerical estimates of $K_\mathrm{Ko} $ from DNS appears to be slightly larger than its theoretical value, which is approximately $1.6$~\cite{Kraichnan:JFM1971_2D3D,Yakhot:JSC1986}.  The increases in the value of $K_\mathrm{Ko} $ in DNS is possibly due to the fact that the inertial range is not completely established, and the the inertial range $E(k)$ is affected by the forcing and dissipation scales even for $4096^3$ grid simulations.  Similar enhancement in the value of $K_\mathrm{Ko}$ has been reported by \citep{Yeung:PRE1997}. \citet{Mininni:PRE2008} observe a  decrease in $K_\mathrm{Ko}$ with the Reynolds number, as in our simulations.   

An  examination of the normalized spectrum $\tilde{E}(\tilde{k})$ indicates a bump  near the transition region between the inertial range and dissipation range ($0.04 \lesssim \tilde{k} \lesssim 0.2$), which is due to the bottleneck effect~\cite{Saddoughi:JFM1994,Falkovich:PF1994,Lohse:PRL1995,Yeung:PRE1997,Dobler:PRE2003,Verma:JPA2007,Donzis:JFM2010Bottleneck}.  The  predicted  $\tilde{E}(\tilde{k})$ by the models of Pao and Pope gradually   decrease with $\tilde{k}$. Thus, these models do not capture the bottleneck effect. This is possibly because Pao's and Pope's models do not address the fluctuations in the energy flux.  
 Nevertheless, the spectrum in the dissipative range is captured reasonably well by these models.


In Fig.~\ref{fig:spectrum_turb}(d, e, f) we plot the nondimensionalized energy fluxes  $\tilde{\Pi}(\tilde{k})$ computed using the DNS data. We observe that  $\tilde{\Pi}(\tilde{k})$ are approximately  constant in the inertial range, consistent with  Kolmogorov's theory~\cite{Kolmogorov:DANS1941Structure}.   In the same plot, we present  the energy fluxes computed using the Pao's and Pope's models (Eqs.~(\ref{eq:Pao_fetak}) and (\ref{eq:Pope_Pik})).  In the inertial range, the predictions of both the models  are in good agreement with the DNS results.   In the dissipation range, the predictions of Pao's model are close  to the numerical values of $\tilde{\Pi}(\tilde{k})$ for $\mathrm{Re}=5.7 \times 10^3$ and $\mathrm{Re}=4 \times 10^{-4}$, but Pao's prediction for $\mathrm{Re}=8 \times 10^{-5}$ is slightly larger than the numerical values. The predictions of Pope's model are systematically lower than their corresponding numerical counterparts. The suppression of the energy flux at the bottleneck region may be a reason for the discrepancy between the predictions of Pao's model and the numerical values.   The increase in $E(k)$ at the bottleneck region leads to an enhanced viscous dissipation, and thus a lower energy flux. As expected, this feature  get more  pronounced at larger Reynolds numbers.

\begin{figure}[htbp]
\begin{center}
\includegraphics[scale = 1.2]{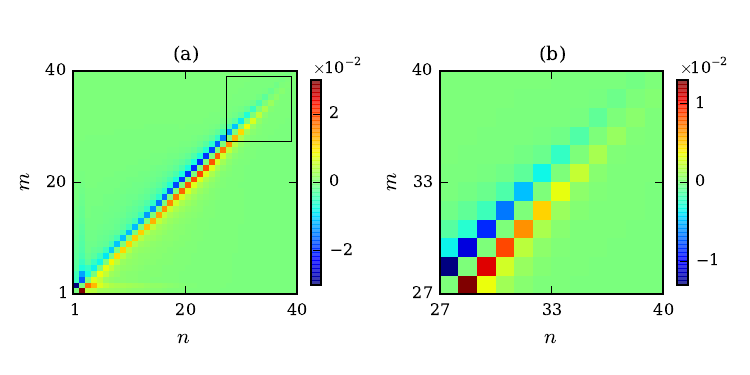}
\end{center}
\setlength{\abovecaptionskip}{0pt}
\caption{For the turbulent simulation on $4096^3$ grid: Plots of the shell-to-shell energy transfer rates (a) for the whole wavenumber range, (b) for the dissipative range corresponding to the boxed region of subfigure (a).   Here $m$ denotes the giver shell, while $n$ denotes the receiver shell.  Our results indicate forward and local energy transfers in the inertial as well as in the dissipative wavenumber range. }
\label{fig:S2S_turbulent}
\end{figure}

In addition, we also study the properties of the shell-to-shell energy transfers for the numerical data of $4096^3$ grid. For this purpose we divide the Fourier space into $40$ shells, whose centers are at the origin ${\bf k}=(0,0,0)$. The inner and outer radii of the shells are $k_{n-1}$ and $k_n$ respectively, where $k_n=\{0, 2, 4, 8 \times 2^{s(n-3)}, ..., 2048\}$  with $s=1/5$. The shells are logarithmically binned~\cite{Debliquy:PP2005}. Note that the $27$th shell, whose wavenumber range is $194 \leq k \leq 223$, separates the dissipative range from the inertial range.  In Fig.~\ref{fig:S2S_turbulent}(a), we exhibit the shell-to-shell energy transfers for the whole range, while Fig.~\ref{fig:S2S_turbulent}(b) shows these transfers for the dissipative range only.  As expected, in the inertial range, shell $m$ transfers energy dominantly to shell $m+1$, and it receives energy from shell $m-1$.  Hence, the shell-to-shell energy transfers are forward and local~\cite{Domaradzki:PF1990,Zhou:PF1993,Verma:Pramana2005S2S}.  Interestingly, similar behaviour, forward and local  energy transfer, is also observed for the wavenumber shells in the dissipative regime.  This is essentially because  the correlations induced by forcing at small wavenumbers are lost deep inside the inertial and dissipative ranges.   

\subsection{Laminar Flow}
\label{subsec:numerical_viscous}

We performed  direct numerical simulation of laminar flows on  $64^3$ grid for four sets of parameters  of  Table~\ref{table:simulation_details_viscous}.  We choose random initial condition for all our simulations. To reach a steady state, we employ random forcing in the wavenumber band $2 \leq k \leq 4$ with a energy supply rate of unity.  The Reynolds numbers of these simulations range from $17.6$ to $49$. For  the steady state, the energy dissipation rate ranges from 0.999 to 1.002, and which is within 0.1-0.2\% of the energy supply rate.

\setlength{\tabcolsep}{48pt}
\begin{table}[htbp]
\begin{center}
\caption{Parameters of our direct numerical simulations (DNS) for laminar flows: kinematic viscosity $\nu$; Reynolds number $\mathrm{Re}$; Kolmogorov's wavenumber $\bar{k}_d$; and $k_{max}\eta$. }
\hspace{3 mm}
\begin{tabular}{c  c  c c}
\hline \hline
$\nu$  & $\mathrm{Re}$   & $\bar{k}_d$ & $k_{max}\eta$ \\[1 mm]
\hline
$0.12$  & $49$ &  $0.9$ & $6.52$ \\
$0.16$  & $32.4$   & $0.7$ & $8.09$ \\
$0.20$  & $23.1$  &$0.6$ & $9.57$ \\
$0.24$  & $17.6$   & $0.5$ & $10.97$ \\
\hline

\end{tabular}
\label{table:simulation_details_viscous}
\end{center}
\end{table}

\begin{figure}[htbp]
\begin{center}
\includegraphics[scale = 0.9]{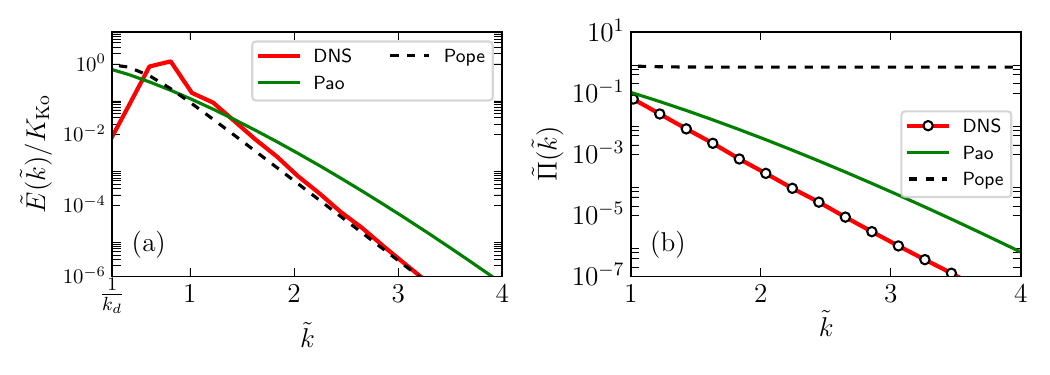}
\end{center}
\setlength{\abovecaptionskip}{0pt}
\caption{For the laminar flow simulation with $\mathrm{Re}=49$: Plots of (a) the normalized energy spectrum $\tilde{E}(\tilde{k})$; (b) the normalized energy flux $\tilde{\Pi}(\tilde{k})$.  See Eqs.~(\ref{eq:Pi_tilde_k}, \ref{eq:E_tilde_k})  for definitions.   The model predictions of Pao (thin line) and Pope (dashed line) do not match with the numerical plots.}
\label{fig:model_dissipation}
\end{figure}

We attempt to verify whether Pope's and/or Pao's models describe the energy spectrum and flux of laminar flows.  Towards this goal, for the laminar flow with $\mathrm{Re} = 49$, in Fig.~\ref{fig:model_dissipation}(a,b) we  plot  the normalized energy spectrum $\tilde{E}(\tilde{k})$ and the normalized energy flux $\tilde{\Pi}(\tilde{k})$.  In the figure, we also plot the predictions of Pao's and Pope's models.  These predictions differ significantly  from the numerical results.  Thus, Pao's and Pope's models do not describe $E(k)$ and $\Pi(k)$ of the laminar flows.    We will show below that the model discussed in Sec.~\ref{sec:model_viscous} describes the numerical results quite well.

\begin{figure}[htbp]
\begin{center}
\includegraphics[scale = 1.1]{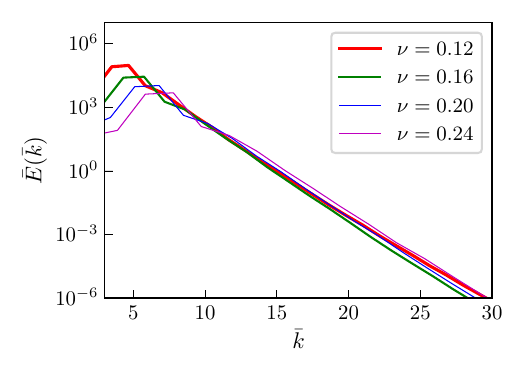}
\end{center}
\setlength{\abovecaptionskip}{0pt}
\caption{For the laminar flow simulations,  plots of the normalized energy spectra of $\tilde{E}(\tilde{k})$ of Eq.~(\ref{eq:Ek_nondim_laminar}).  All the plots {merge} into a single curve for $k> k_f$.}
\label{fig:spectrum_laminar}
\end{figure}

\begin{figure}[htbp]
\begin{center}
\includegraphics[scale = 1.1]{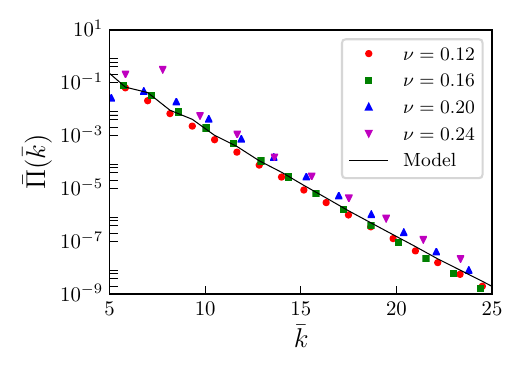}
\end{center}
\setlength{\abovecaptionskip}{0pt}
\caption{For the laminar flow simulations,  plots of the normalized energy fluxes. All the plots {merge} into a single curve.}
\label{fig:flux_dissipation}
\end{figure}

 In Fig.~\ref{fig:spectrum_laminar}, we plot $\bar{E}(\bar{k}) = \mathrm{Re}^3 E(k)k/u_\mathrm{rms}^2$ (see Eq.~(\ref{eq:model_laminar_Ek}))  computed using the numerical data for $\mathrm{Re} = 49, 32.4, 23.1$ and 17.6. We find that for $k>k_f$, all $\bar{E}(\bar{k})$'s {merge} into a single curve  indicating that   $\bar{E}(\bar{k})$ is a universal function in this range. Also, $\bar{E}(\bar{k}) \sim \exp(-\bar{k})/k$ verifying  the model predictions (see Sec.~\ref{sec:model_viscous}).     Note that the $\bar{E}(\bar{k}) $ for low $k$ does not {merge} into a single curve.

In Fig.~\ref{fig:flux_dissipation}, we  plot the normalized energy flux  $\bar \Pi (\bar k)$ for the  four simulations. We observe that the  function $ k \exp(-\bar{k})$ provides a good fit to  the numerical  $\bar \Pi (\bar k)$, consistent with the model predictions.   The aforementioned consistency between the numerical results and  model predictions yields strong credence to the model.   

\begin{figure}[htbp]
\begin{center}
\includegraphics[scale = 1.2]{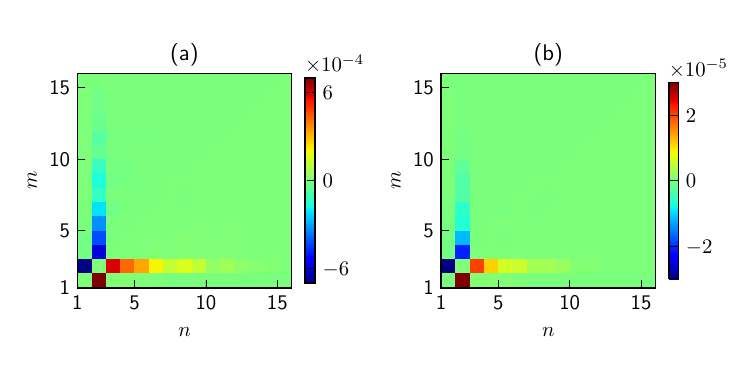}
\end{center}
\setlength{\abovecaptionskip}{0pt}
\caption{Plots of the shell-to-shell energy transfer rates for the laminar simulation with  (a)  $\mathrm{Re}=49$, (b) $\mathrm{Re}={17.6}$.   Here $m$ denotes the giver shell, while $n$ denotes the receiver shell.  The forcing wavenumbers belong to the 2nd shell.}
\label{fig:S2S_laminar}
\end{figure}

We compute the shell-to-shell energy transfers using the numerical data for  $\mathrm{Re} = 49$ and $17.6$.  We divide the Fourier space into $32$ shells, whose centers are at the origin ${\bf k}=(0,0,0)$. The inner and outer radii of the shells are $k_{n-1}$ and $k_n$ respectively, where $k_n=\{0, 2, 4, 8, 8 \times 2^{s(n-3)}, ..., 32\}$ with $s=1/27$.  The forcing wavenumber band $2 \leq k \leq 4$ is inside the 2nd shell.    In Fig.~\ref{fig:S2S_laminar}(a,b), we exhibit the shell-to-shell energy transfers for $\mathrm{Re}=49$ and $17.6$ respectively.  We observe that the most dominant energy transfers are from the forcing band to the shells of larger radii, e.g., from shell 2 to  shells 3-10 for $\mathrm{Re}=49$, and to the shells 3-7 for $\mathrm{Re}=17.6$.  Thus, the energy transfers for laminar flows are {\em nonlocal}.  This is because in laminar flows, the velocity field appears to be correlated with the forcing field.   This issue needs further investigation.

\section{Conclusions}
\label{sec:conclusions}

Turbulence is a complex problem, hence we rely on turbulence models.  Pao~\cite{Pao:PF1965} and Pope~\cite{Pope:book} constructed turbulence models that explains the turbulence behaviour in the inertial and dissipative ranges.  Several experimental results on the energy spectrum have been compared with the model predictions, and they match with each other quite well. To best of our knowledge, ours is the first numerical verification of Pao's and Pope's models~\cite{Pao:PF1965,Pope:book}.  The present paper shows that the predictions of the above models  and numerical results are consistent with each other, except minor differences in the energy flux in the dissipation regime.

The aforementioned models for turbulent flows, however, have certain deficiencies.  The hump in the energy spectrum near the beginning of dissipation range is related to the bottleneck effect~\cite{Falkovich:PRL2005,Verma:JPA2007,Donzis:JFM2010Bottleneck}; this hump is not captured by  of Pao's and Pope's models.  Also, the numerical values of the energy flux in the dissipative regime differ from the model predictions by a small amount.  Thus, the models of Pao~\cite{Pao:PF1965} and Pope~\cite{Pope:book} need to be revised.  It is also interesting to note that Pao's model~\cite{Pao:PF1965} does not involve any free parameter (except Kolmogorov's constant $K_\mathrm{Ko}$) in comparison to several free parameters in Pope's model.  The parameters of Pope's model are chosen so as to fit with $E(k)$ derived from experiments.  We show that for turbulent flows, the shell-to-shell energy transfers are forward and local in both inertial and dissipative ranges.

In this paper, we also present a new model for the energy spectrum and flux of laminar flows with $\mathrm{Re}\sim 1$.  According to our model, the energy spectrum and flux exhibit exponential behaviour $(\exp(-k))$.  We verify the model predictions using numerical simulations.  For the laminar flows, we also show that the energy transfers are nonlocal and forward; the forcing wavenumbers supply energy to different shells.  For moderate $\mathrm{Re}$ ($\sim 25$) to  large  $\mathrm{Re}$, Mart\'{i}nez {\em et al.}~\cite{Martinez:JPP1997}  argued that the energy spectrum  is of the form Eq.~(\ref{eq:Ek_Martinez}), where the parameters $\alpha$ and $\beta$ depend on the Reynolds numbers and length scales.   Our model for the laminar flow is simpler and more suitable than that by Mart\'{i}nez {\em et al.}~\cite{Martinez:JPP1997}.

It is important to differentiate the behaviour of laminar flows with that of highly viscous flows for which  $\mathrm{Re} \rightarrow 0$.  For the highly viscous flows, the nonlinear term vanishes, and the velocity field is computed using $\nu \nabla^2 {\bf u} = -{\bf f}$, or ${\bf u(k)} = {\bf f(k)}/(\nu k^2)$ in Fourier space. In such flows, the energy flux vanishes due to the absence of the nonlinear term. An injection of weak nonlinearity in a highly viscous flow will induce a small flux that can be computed perturbatively.

In summary, in this paper, we verify the predictions of Pao's and Pope's models~\cite{Pao:PF1965,Pope:book} for turbulent flows.  We also show that the energy spectrum and flux of laminar flows are of the form $\exp(-k)$.

\section*{Acknowledgements}
We thank Mohammad Anas for a valuable feedback on $\bar{k}_d$. Our numerical simulations were performed on Cray XC40 {\em Shaheen} II at KAUST supercomputing laboratory, Saudi Arabia and {\em Chaos} cluster of IIT Kanpur. This work was supported by the research grants PLANEX/PHY/2015239 from Indian Space Research Organisation India, INT/RUS/RSF/P-03 by the  Department of Science and Technology India, and RSF-16-41-02012 by Russian Science Foundation for the Indo-Russian project.



%

\end{document}